\documentclass{article}
\usepackage{waspaa23,amsmath,graphicx,url,times}
\usepackage{color}
\usepackage{hyperref}
\usepackage{csquotes}
\usepackage[nameinlink]{cleveref}
\usepackage[ruled]{algorithm2e}
\usepackage{mathtools}
\usepackage{xparse}
\usepackage{amsfonts}
\usepackage{tabularx}
\usepackage{pifont}
\usepackage{amssymb}
\newcommand*{\ditto}{---\texttt{"}---}

\NewDocumentCommand{\ceil}{s O{} m}{%
  \IfBooleanTF{#1} 
    {\left\lceil#3\right\rceil} 
    {#2\lceil#3#2\rceil} 
}
\DeclareMathOperator*{\argmin}{argmin} 
\crefname{algocf}{alg.}{algs.}
\Crefname{algocf}{Algorithm}{Algorithms}
\usepackage[symbol]{footmisc}
\title{\vspace{-1.0cm}Compressing audio CNNs with graph centrality based filter pruning}
\name{\vspace{-0.2cm}James A King$^{+}$,
      Arshdeep Singh$^{+}$,
      Mark D. Plumbley,
      }
      
\address{Centre for Vision, Speech and Signal Processing (CVSSP), \\ University of Surrey, UK.}
\begin{document}

\ninept
\maketitle
\def\thefootnote{+}
\footnotetext{Both authors contributed equally.}
\def\thefootnote{\arabic{footnote}}
\begin{sloppy}

\begin{abstract}

Convolutional neural networks (CNNs) are commonplace in high-performing solutions to many real-world problems, such as audio classification. CNNs have many parameters and filters, with some having a larger impact on the performance than others. This means that networks may contain many unnecessary filters, increasing a CNN's computation and memory requirements while providing limited performance benefits. To make CNNs more efficient, we propose a pruning framework that eliminates filters with the highest \enquote{commonality}. We measure this commonality using the graph-theoretic concept of \textit{centrality}.
We hypothesise that a filter with a high centrality should be eliminated as it represents commonality and  can be replaced by other filters without affecting the performance of a network much. An experimental evaluation of the proposed framework is performed on acoustic scene classification and audio tagging. On the DCASE 2021 Task 1A baseline network, our proposed method reduces computations per inference by 71\% with 50\% fewer parameters at less than a two percentage point drop in accuracy compared to the original network. For large-scale CNNs such as PANNs designed for audio tagging, our method reduces  24\% computations per inference with 41\% fewer parameters at a slight improvement in performance.

\end{abstract}


\section{Introduction}
\label{sec:intro}
Convolutional neural networks (CNNs) have shown promising results in a variety of audio tasks, including speech recognition \cite{tu2019speech}, music analysis \cite{yu2020deep} and audio classification \cite{purwins2019deep, kong2020panns}. Typically, CNNs have many layers, such as convolutional and pooling layers. Each layer consists of parameters, including weights, biases and filters, which are all learned through an optimisation process for the given problem. CNNs often have many parameters, requiring a large amount of memory storage. Moreover, the majority of computations during inference are performed by convolution operations, which involve sliding a set of filters over the input data to create a feature map; this computation is especially time-consuming when dealing with large input data and a large number of filters. While CNNs are highly effective in solving non-linear complex tasks \cite{denton2014exploiting}, the requirement of high memory and heavy computations during inference is a bottleneck to deploying them on resource-constrained devices such as mobile phones or Internet of things (IoT) devices \cite{Heittola2020}. Moreover, the prolonged need for large computations in machine learning (ML) models contributes heavily to $\text{CO}_2$ emissions, making larger CNNs environmentally unfriendly. For instance, a modern GPU (e.g. NVIDIA GPU RTX-2080 Ti) used to train ML models for 48 hours generates the equivalent $\text{CO}_2$ emitted by an average car driven for 13 miles\footnote{\href{https://mlco2.github.io/impact/\#\#compute}{Machine learning $\text{CO}_2$ estimator}}. Thus, the issue of reducing the computations and memory requirement for CNNs has drawn a significant amount of attention in the research community.

Recent efforts towards compressing CNNs involve filter pruning methods \cite{liang2021pruning}, which eliminate unimportant filters which contribute the least to performance. These methods measure the importance of the filters using either active or passive methods. Active methods \cite{lin2020hrank, yeom2021toward} use a dataset to generate feature maps from the filters and then measure the importance of the filters using various measures such as entropy, rank or the average percentage of zeros on feature maps. Some active methods even identify important filters during the training of CNNs by involving extra parameters such as soft mask to each filter and then jointly optimise the CNN parameters and soft mask \cite{he2018soft,lin2019towards}. Conversely, passive methods are data-free, using only the filters to quantify their own importance. Therefore, passive methods are easier to apply and require significantly less storage than active filter pruning, something particularly important for larger models. 

Passive filter pruning methods are either norm-based \cite{li2016pruning}, which computes $l_1$ or $l_2$ norm of the filters to define their importance, or similarity-based \cite{singh2022passive}, where similar filters are removed. Norm-based methods are based on a smaller-norm-less-important criterion and eliminate filters with the smaller norm. However, eliminating smaller norm filters may ignore the diversity learned in the network and redundancy in the high-norm filters. Similarity-based pruning methods capture this diversity, eliminating redundant filters based on the pairwise similarity between filters \cite{singh2022passive}. Such similarity-based methods give better performance compared to norm-based methods. However, the pairwise similarity method eliminates redundant filters by considering only the similarity between pairs, where the closest filters might differ. Ignoring such filters may reduce the useful diverse information learned in the network.

In this paper, we propose a passive filter pruning method where a filter is considered redundant if others can replace it. To measure its redundancy, we consider filters as nodes in a graph and determine the centrality of each node. A high node centrality represents a node with high commonality among any other two nodes. By ranking the centrality, we better understand the effect removing such common filters would have in the network over the previous method \cite{singh2022passive}, where commonality is measured only within the closest pairs of filters without considering commonality with other filters.






The rest of this paper is organised as follows. \Cref{sec:back}
presents a background of pruning methods. The proposed method to identify similar filters is described in \Cref{sec:meth}. Next, \Cref{sec:exp} includes the experimental setup. The results and analysis are included in Section \Cref{sec:perf}. Finally, the discussion and conclusions are presented in \Cref{sec: disc} and \Cref{sec: concl}, respectively.
\section{Background and Related work}
\label{sec:back}
\subsection{The Pruning problem}
The filter pruning problem is one of the main ways to reduce a network size by eliminating some filters while maintaining performance. Given a CNN with $L$ convolutional layers, each with a set of filters $\mathcal{F} = \{F_1, F_2, \dots, F_n\}$, the aim to find a pruned CNN having layer $L^{\prime}$ with a reduced set of filters $\mathcal{F}^{\prime} \subseteq \mathcal{F}$ such that $p\%$ of filters are removed.

\begin{equation}
\begin{aligned}
    & \min_{\mathcal{F}^{\prime} \subseteq \mathcal{F}, |\mathcal{F}^{\prime}| = \ceil[\big]{(1-p)|\mathcal{F}|}} \mathcal{C}(L^{\prime}) 
    &\text{s.t.} \quad \mathcal{P}(L^{\prime}) \gtrapprox (1 - \epsilon) \mathcal{P}(L)
\end{aligned}
\end{equation}
where $\epsilon$ is a small tolerance value.
So $|\mathcal{F}^{\prime}| =  \ceil[\big]{(1-p)|\mathcal{F}|}$, and the performance metric $\mathcal{P}(L^{\prime})$ remains close to $\mathcal{P}(L)$. It is also not uncommon to find $\epsilon$ having a negative value, so pruning improves performance.

To select a few important filters, $\mathcal{F}^{\prime}$,  the importance of the filters is computed using active or passive filter pruning methods. After obtaining the pruned network, a fine-tuning process is performed to regain most of the performance. This fine-tuning process re-trains the pruned network on the original data.

\subsection{Methods to compute CNN filter importance}

\textbf{Active methods:} Active filter pruning methods are data-driven, which allows the evaluation of $\mathcal{P}(L^{\prime})$ to influence which filters are pruned or retained. For example, previous attempts compute the importance of filters during the training process by jointly learning the CNN's and the extra parameters, such as the soft mask associated with each of the CNN parameters \cite{lin2019towards,liu2017learning}.
However, these methods add up to 10 times more training time \cite{lagunas2021block,xia2022structured} and are computationally expensive. Other active filter pruning methods generate features corresponding to a set of examples and then apply metrics such as rank \cite{lin2020hrank}, energy \cite{yeom2021toward}, the average percentage of zeros \cite{hu2016network} to quantify the importance of filters or similarity measures such as clustering \cite{wu2022filter} on feature maps to eliminate filters corresponding to redundant feature maps. However, these methods use extra memory resources to obtain feature maps and involve complex training procedures to optimize extra parameters such as a soft mask, particularly when a large-scale pre-trained network is used for downstream tasks.


\noindent \textbf{Passive methods:} Passive approaches are data-free and do not evaluate $\mathcal{P}(L^{\prime})$ during the filter selection process. Therefore, the passive approach is much more scalable and efficient than the active methods. For example, Li et al. \cite{li2016pruning} compute the $l_1$-norm, the absolute sum of the filter parameters, of the filters to quantify filter importance and eliminate low-norm filters to obtain the pruned network. He et al. \cite{he2019filter} computes the geometric median of the filters and eliminates the redundant filter which are closer to the geometric median of all filters. However, previous methods are based on the smaller-norm-less-important criterion where a filter is considered less important if the filter has low $l_1$/$l_2$-norm from the origin or from the geometric median of the filters, ignoring the redundancy in the filters with high norm. Moreover, the diversity in selecting filters is also ignored as only high-norm filters would be considered as important. Other methods capture the diversity using similarity-based measures. For example, Kim et al. \cite{park2020reprune} perform clustering on filters, select a filter from each cluster as important, and eliminate the other filters. Singh et al. \cite{singh2022passive} measure similarity between filters by computing a pairwise cosine distance for all filters and then eliminating a filter from a pair of similar filters.

\subsection{Optimal Pruning Heuristic}

We need a good heuristic indicating the solution's quality when solving the passive pruning problem. We use the heuristic that the filters removed should be the least similar to those kept so we resolve the most redundancy.
To get the optimal set of filters $\mathcal{F}_\text{pruned}$ with $|\mathcal{F}_\text{pruned}|=\ceil[\big]{(1-p)|\mathcal{F}|}$:

\begin{equation}
\label{eq:optimal}
    \mathcal{F}_\text{pruned} = \argmin_{\mathcal{F}^{\prime} \subseteq \mathcal{F}, |\mathcal{F}^{\prime}| = \ceil[\big]{(1-p)|\mathcal{F}|}} \sum\limits_{i=1}\sum\limits_{j=1} \textit{sim}(\mathcal{F}^{\prime})_{i,j}
\end{equation}
where \textit{sim} measures how similar two filters are. In this way, we have modelled the total present similarity in a matrix. Implementing an algorithm to calculate this function precisely would require iterating over all possible subsets of $\mathcal{F}$ and so unfeasible for a relatively small number of filters.

\begin{figure}[t]
\centering
  \includegraphics[scale =0.34]{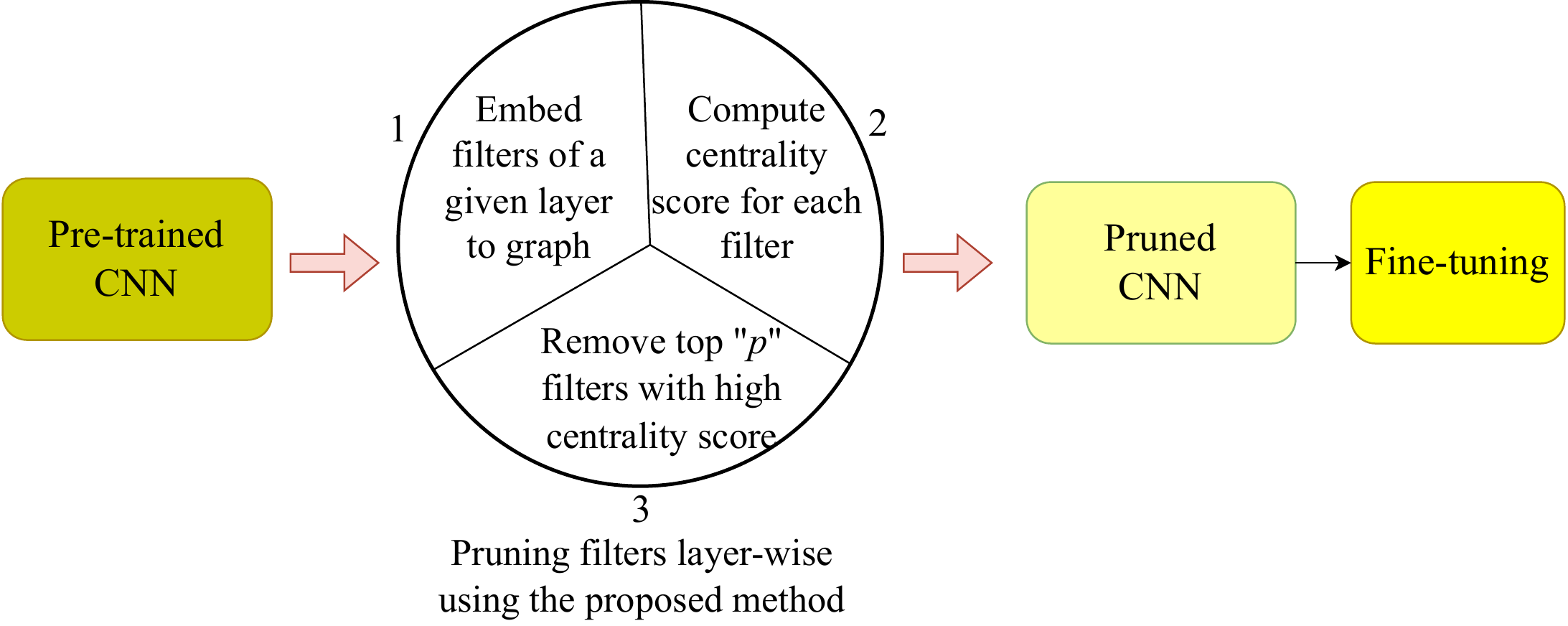}
  \vspace{-0.45cm}
  \caption{An overall pipeline of the proposed framework. A pre-trained CNN is pruned layer-wise using centrality measures, followed by a fine-tuning process to regain most of the performance.}
  \label{fig: overall}
\end{figure}



\section{Proposed Method}
\label{sec:meth}
The proposed method prunes a CNN layer-by-layer. Each layer contains a set of $n$ filters $\mathcal{F}$ where each $F\in\mathcal{F}$ is a 3D tensor of shape ($w \times h \times c$). We start by flattening each filter $F$ into $F^\text{flat}$ of shape ($wh \times c$) with $F^\text{flat}_{(i-1)h + j, k} = F_{i, j, k}$.  We then perform singular value decomposition (SVD) on each flattened filter $F^\text{flat}$ to find its best Rank-1 approximation under the Frobenius norm; this yields the most significant singular value $\sigma_1$ and corresponding left and right singular vector $l_1,r_1$. We can now approximate the original filter with $\hat{F} = \frac{{(\sigma_1 l_1 r_1^T)}_{i,j}}{\sqrt{\sum_{k=1}^{wh} ({(\sigma_1 l_1 r_1^T)}_{k,j})^2}}$. Then we pick any column from $\hat{F}$ as our filter representative $f$. We can pick any column because our Rank-1 approximation means they will all be identical. Finally, we compute the cosine similarity between each filter representative $f_n$ in the layer giving us a similarity matrix of shape $\mathbf{W}\in \mathbb{R}^{n\times n}$. Utilizing $\mathbf{W}$, we perform following steps to obtain a importance scores for a given layer. After this pre-precessing is complete, there are three other steps taken.

\noindent \textbf{1. Embed filters to graph:}
We construct a graph $G = (\mathcal{F}, E)$ where $\mathcal{F}$ is the set of vertices representing filters, $E$ is the set of edges with property $w(e)=\mathbf{W}_{u,v}$ that assigns a weight to each edge $e=(u,v) \in E$. Let $S_N(G)$ be the set of all complete subgraphs of $G$ of size $N$. so that:
\begin{multline}
  S_N(G) = \{H \mid H \subseteq G, |\mathcal{F}_H| = N, \text{ and } \\E_H = \{ (u,v) \in E \mid u \in \mathcal{F}_H \text{ and } v \in \mathcal{F}_H \} \}
\end{multline}
So for $H \in S_{\ceil[\big]{(1-p)|\mathcal{F}|}}(G)$, $H = (\mathcal{F}_H, E_H)$ is a subgraph of $G$ with $|\mathcal{F}_H| = \ceil[\big]{(1-p)|\mathcal{F}| }$.
The subgraph $H_{pruned}$ can hence be defined:
\begin{equation}
\label{eq:graph}
H_{\text{pruned}} = (\mathcal{F}_\text{pruned},E_\text{pruned}) = \text{argmin}_{H \in S_{\ceil[\big]{(1-p)|\mathcal{F}| }}(G)} \sum_{e \in E_H} w(e)
\end{equation}
Where the computation of $\mathcal{F}_\text{pruned}$ is equivalent to \Cref{eq:optimal}.

\noindent \textbf{2. Compute centrality scores for each filter:} We utilise this graphical setup and two centrality algorithms to compute the importance of each filter.

\textbf{Weighted degree centrality (WDC)\cite{Singh2020Centdefs}:} Our first approach uses weighted degree centrality (WDC) to assign importance to each node. Then we keep the filters corresponding to the $\ceil[\big]{(1-p)|\mathcal{F}|}$ lowest scoring nodes since the highest scoring filter is similar to most other filters. We compute this importance with
\begin{equation}
    C(V) = \sum_{e \in \text{Edges of V}} w(e)
\end{equation}
and so approximate \Cref{eq:graph} with 
\begin{multline}
\label{eq:aprox}
\mathcal{F}_{pruned} = 
    \{F_i \in \mathcal{F} \mid v_i \text{ is one of the } \ceil[\big]{(1-p)|\mathcal{F}| } \\\text{ largest elements in } \{C(F) \mid F \in \mathcal{F}\}\}\\
\end{multline}

This equation is much easier to solve and can trivially be done so in Polynomial-time approximation, making it feasible to calculate with minimal performance loss.

\textbf{Betweenness centrality (BC)\cite{Singh2020Centdefs}:} We also explore the idea of using betweenness centrality (BC) to perform pruning. BC gives each node in a network a score based on how important it is to the network's connectivity. We quantise this by measuring the shortest path between each node and counting how many times each node is in any such path. Therefore, the more minimum paths that go via a node, the more central a node is, and the more likely it is that its removal will decrease the total similarity of the network. To perform these experiments, we have the same approximator as \Cref{eq:aprox} but we score nodes with

\begin{equation*}
    C(v) = \sum_{s,t \in V} \frac{\sigma(s,t|v)}{\sigma(s,t)}
\end{equation*}
where $\sigma(s,t)$ is the total number of shortest paths from node $s$ to node $t$, and $\sigma(s,t|v)$ is the number of those shortest paths that pass through the node $v$. The sum is taken over all pairs of nodes $s$ and $t$ in the graph.

\noindent \textbf{3. Obtaining Pruned network and  fine-tuning:} After obtaining centrality scores for each filter in a given layer, we prune $p$ filters with high centrality scores in that layer. Then, we repeat the same procedure, steps (1) and (2), for other layers as well. After removing the filters from various convolutional layers a pruned network is obtained. In the end, the pruned network is retrained to regain most of the lost performance due to elimination of some filters. \Cref{fig: overall} shows an overall flow of the proposed method.

\section{Experimental setup}
\label{sec:exp}

We evaluate the proposed pruning framework on CNNs designed for acoustic scene classification (ASC) and audio tagging. An overview of the unpruned CNNs is given below:

\textbf{(a) DCASE21\_Net:} We use a publicly available DCASE 2021 Task 1A baseline network designed for ASC to classify 10 different acoustic scenes \cite{martin2021low} and denote it as \enquote{DCASE21\_Net}. DCASE21\_Net consists of three convolutional layers (termed as C1 to C3) and one fully connected layer. The network takes input a log-mel spectrogram of size (40 $\times$ 500), corresponding to a 10 second audio clip  and is trained with the Adam optimizer for 200 iterations. The network has 46,246 parameters and requires approximately $287$M multiply-accumulate operations (MACs) during inference per input, and gives 48.58\% accuracy.

\textbf{(b) PANNs\_CNN14:} PANNs \cite{kong2020panns} are large-scale pre-trained audio neural networks designed for audio tagging. PANNs are trained on the AudioSet \cite{gemmeke2017audio}, which contains over 2M labelled sound events comprising 527 different sound classes. For our experiments, we use one of the PANNs models, CNN14, which consists of 12 convolutional layers (denoted as C1 to C12) and denote the network as \enquote{PANNs\_CNN14}.

PANNs\_CNN14 takes a log-mel spectrogram of size (1000 $\times$ 64) as an input. CNN14 gives 0.431 mean average precision (mAPs) and 0.973 area under the curve (AUC) for the AudioSet evaluation dataset. CNN14 has 81M parameters and 21G MACs\footnote{\href{https://pypi.org/project/thop/}{MACs computation Pytorch package.}} corresponding to a 10-second-length audio clip sampled at a 32KHz  with a window size of 1024 samples and a hop size of 320 samples. PANNs\_CNN14 is trained with data augmentation techniques such as Mixup and SpecAugment for 600k iterations.

\noindent \textbf{Pruning and fine-tuning:} After obtaining the importance  filters across various convolutional layers using the proposed centrality based pruning method, 
 we eliminate $p \in \{25\%, 50\%, 75\%\}$  top unimportant filters from a subset of convolutional layers to obtain pruned networks. Once the pruned network is obtained, we perform fine-tuning of the pruned network with similar conditions such as loss function, batch size except for fewer iterations as used while training the unpruned network.

For DCASE21\_Net, we consider all convolutional layers for pruning and perform fine-tuning for 100 iterations. For PANNs\_CNN14, we provide a preliminary analysis and fine-tuned the pruned network for 180k iterations by pruning  only C7 to C12 layers as these layers contain approximately 99\% of the parameters.

\noindent \textbf{Other methods for comparison:} The proposed pruning method is compared with methods, (a) $l_1$-norm \cite{li2016pruning}, (b) geometric median (GM) method \cite{he2019filter} and (c) pair-wise similarity method (CS) \cite{singh2022passive}. We also use the active filter pruning methods, including HRank \cite{lin2020hrank} and Energy-aware pruning \cite{yeom2021toward} that uses feature maps for pruning. For pruning, we randomly select 500 training examples to generate feature maps corresponding to each filters. Subsequently, we follow same fine-tuning process as used in the other methods.

\section{Performance analysis}

\label{sec:perf}

\begin{figure}[h]
    \centering\includegraphics[scale=0.4]{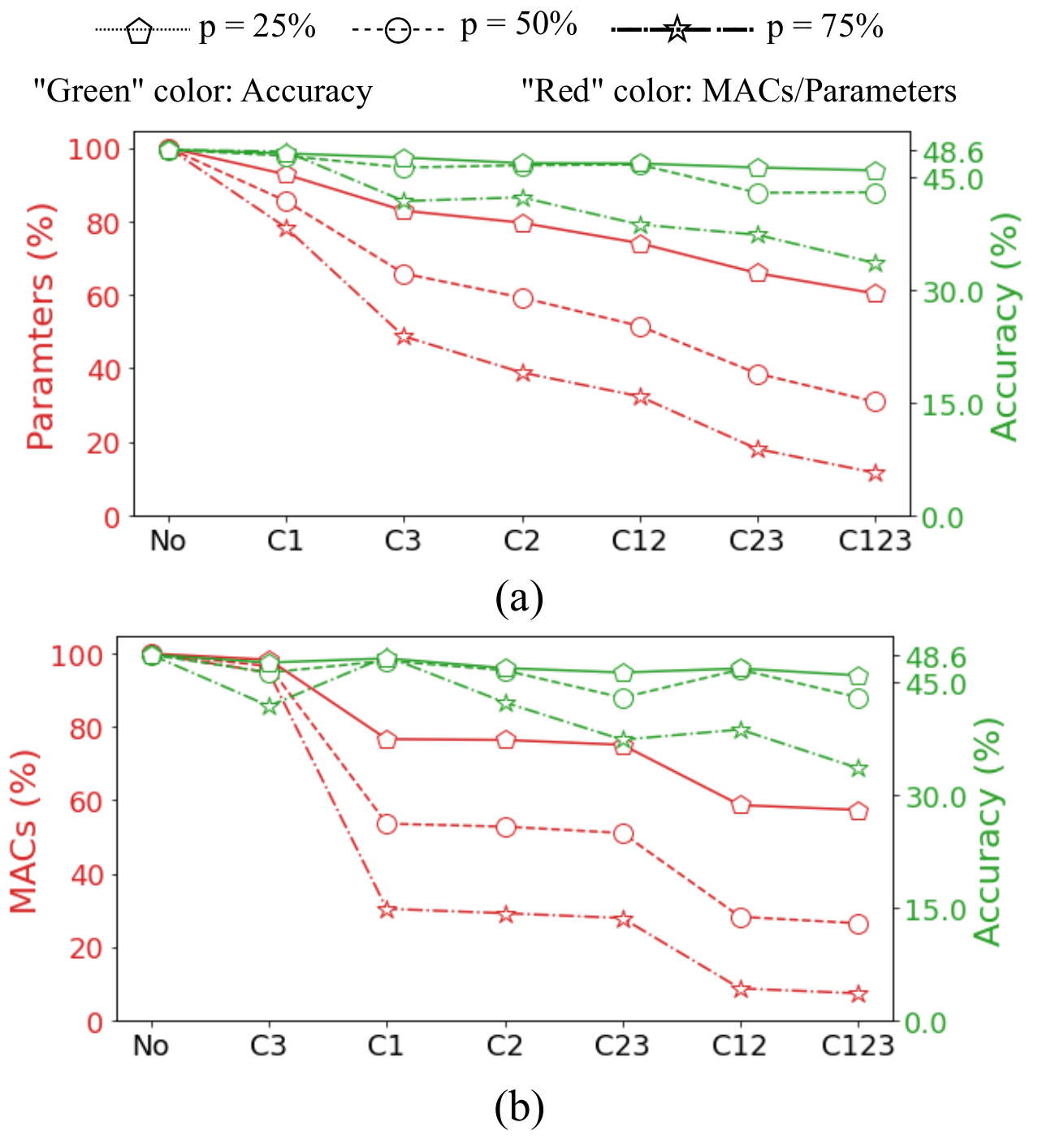}
    \vspace{-0.6cm}
    \caption{Accuracy obtained after pruning various intermediate convolutional layers using betweenness centrality (BC) measure in DCASE21\_Net at different pruning ratios. (a) shows accuracy versus parameters and (b) shows accuracy versus MACs.}
    \label{fig: DCASE21NET}
\end{figure}

\noindent \textbf{DCASE21\_Net:} \Cref{fig: DCASE21NET} shows accuracy, the number of parameters and the number of MACs obtained after pruning various subsets of convolutional layers at different $p$ using BC measure for DCASE21\_Net. Pruning 25\% filters from C3 layer reduces both the number parameters and MACs by 20\% at approximately 1 percentage point drop in accuracy. We find that pruning 25\% filters across various subset of layers result in an accuracy drop of less than 3 percentage points compared to the unpruned network at approximately 60\% reduction in parameters and 40\% reduction in MACs.
As the number of filters pruned across various layers increases from 25\% to 75\%, the accuracy drop, the number of reduced parameters and MACs for the pruned network increases except for C1 layer which shows the least sensitivity towards accuracy drop and the number of MACs at different $p$.

Next, we compare the accuracy obtained using various pruning frameworks in \Cref{tab: comparison...}. For a fair comparison, we obtain the pruned network at the pruning ratio as obtained using the pair-wise similarity method \cite{singh2022passive}. We find that the accuracy obtained using the centrality methods is equal to or greater than that obtained using $l_1$-norm, GM and pairwise-similarity methods. The proposed WDC pruning method  gives approximately similar accuracy without using much memory and 500 examples during pruning process compared to that of the  existing active pruning methods.


\begin{table}[t]
     \caption{DCASE21\_Net comparison with other pruning methods.}
         \centering
         \Large
    \resizebox{0.39\textwidth}{!}{\begin{tabular}{|c|c|c|c|c|c|}
        \hline
       {Pruning Method} & {\begin{tabular}[c]{@{}c@{}}Active\end{tabular}}  &{\begin{tabular}[c]{@{}c@{}}Filter or \\feature map\\ storage\end{tabular}} & {Accuracy(\%)}  & {Parameters} & {MACs}  \\ \hline
       {Baseline(No Pruning)}& {-}& {-} & {48.58} & {46246} & {287M} \\ \hline
             {HRank \cite{lin2020hrank}}& {\ding{51}}& {1.26GB} & {47.24} & {24056} & {139M} \\ 
            {Energy-aware \cite{yeom2021toward}}& {\ding{51}}  & {1.26GB}& {47.00} & {\ditto} & {\ditto}\\ \hline
             {$l_1$-norm \cite{li2016pruning}} & {\ding{53}} & {0.15MB} & {44.42}  & {\ditto} & {\ditto}\\ 
            
            {Similarity-based \cite{singh2022passive}} & {\ding{53}}  & {0.15MB} & {45.54}  & {\ditto} & {\ditto}\\ 
            {GM \cite{he2019filter}} & {\ding{53}}  & {0.15MB} & {45.84}  & {\ditto} & {\ditto}\\ 
             {Proposed (BC)} & {\ding{53}} & {0.15MB} & {45.84}  & {\ditto} & {\ditto}\\ 
             {Proposed (WDC)} & {\ding{53}} & {0.15MB} & {46.91}  & {\ditto} & {\ditto}\\ \hline
    \end{tabular}}
    \label{tab: comparison...}
\end{table}

\noindent \textbf{PANNs\_CNN14:} \Cref{fig: PANNs: maps vs iter} shows mAPs obtained during fine-tuning of the pruned network at different pruning ratios using BC measure. We find that pruning 25\% filters across C7 to C12 layers reduce 41\% parameters  and 24\% MACs at a slight improvement in performance with 0.434 mAPs and 0.974 AUC compared to that of the unpruned network. 
Pruning 50\% filters, the mAPs is 0.426, and the AUC is 0.974 with 70\% fewer parameters and 36\% fewer MACs. Pruning 75\% filters, the mAPs is 0.399, and the AUC is 0.973 with 78\% fewer parameters and 46\% fewer MACs. 



\begin{figure}[]
    \centering
    \includegraphics[scale = 0.23]{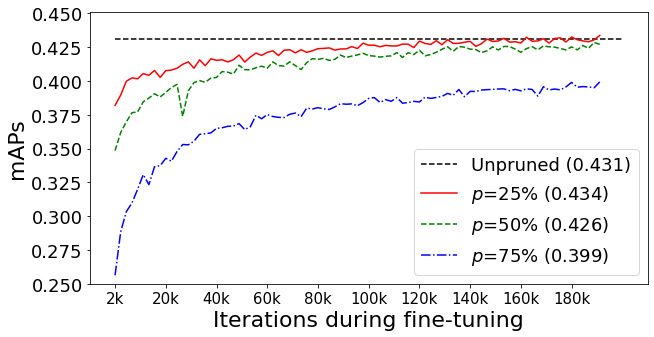}
    \vspace{-0.55cm}
    \caption{mAPs obtained during fine-tuning of the pruned PANNs\_CNN14 at different $p$ using betweenness centrality (BC). Also, the maximum mAPs obtained during the fine-tuning process is  shown in round brackets.}
    \label{fig: PANNs: maps vs iter}
\end{figure}


\begin{figure}[]
    \centering
    \includegraphics[scale = 0.4]{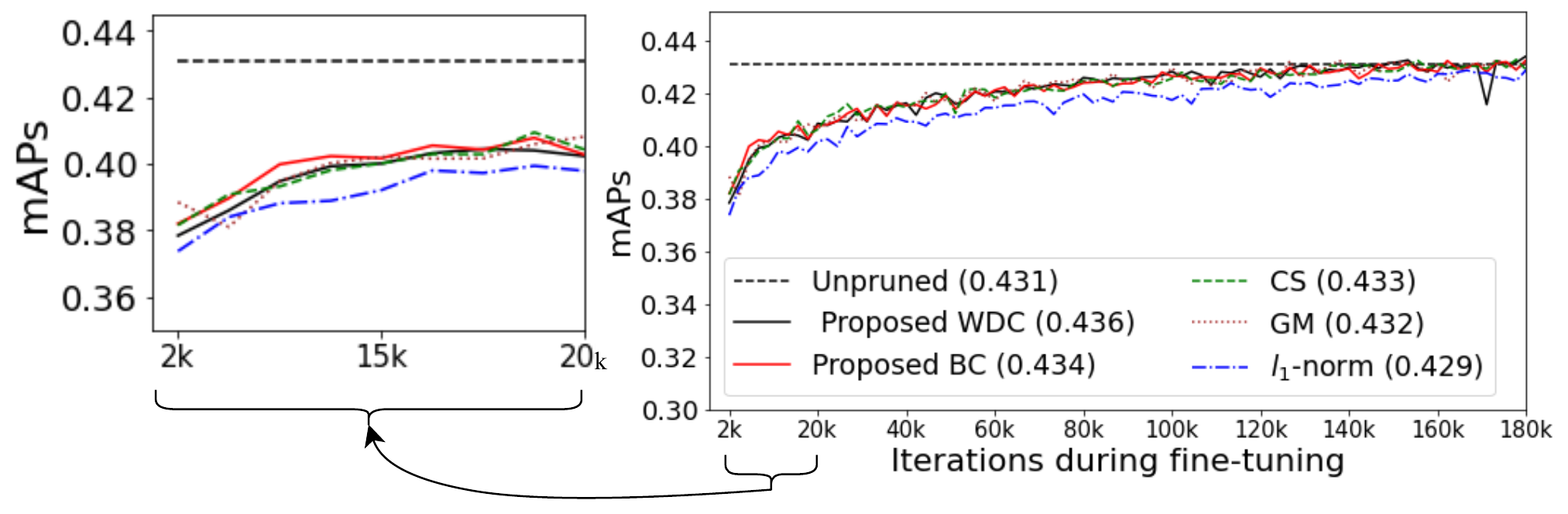}
    \vspace{-0.55cm}
    \caption{mAPs comparison for PANNs\_CNN14 with other pruning methods. Also, the maximum mAPs obtained for each method is shown in round brackets.}
    \label{fig: PANNs: comparison}
\end{figure}

Next, we compare the proposed pruning methods with that of the existing passive pruning methods in \Cref{fig: PANNs: comparison} when  25\% of the filters are pruned across C7 to C12 layers. The  proposed method gives better mAPs  compared to other methods when fine-tuning iterations are less than 15k. After fine-tuning the pruned network for 180k iterations, the proposed method gives slightly improved performance compared to other methods. Overall,  the proposed WDC pruning method results in a pruned network which performs better than the unpruned PANNs\_CNN14 with an advantage of reduced number of MACs and the parameters as well.

\section{Discussion}
\label{sec: disc}
In this paper, we use the graph centrality of filters in a CNN to define their redundancy and pruning. We find that the proposed weighted degree centrality based passive filter pruning method performs better than the existing pairwise-similarity method and norm-based methods. For DCASE21\_Net, our experiments reveal that the proposed passive pruning method achieves similar accuracy compared
to that of the active filter pruning methods without involving any feature maps. For PANNs\_CNN14, we find that the pruned network gives a  slightly better performance compared to that of the unpruned network with approximately 3 times fewer iterations as used in training the unpruned network. This suggests that the existing large-scale pre-trained network can be used efficiently by first applying the proposed passive filter pruning to obtain a smaller-size pruned network, and then perform fine-tuning for few iterations less than that required for unpruned network to achieve similar performance. Hence, the underlying computational resources can be used effectively.

\section{Conclusion}
\label{sec: concl}
This paper presents a passive filter pruning method to reduce the computational complexity and memory storage of CNNs by exploring the graph centrality of the filters. The proposed pruning method achieves similar or better performance compared to that of existing norm-based and pairwise similarity methods, showing the advantage of utilising graph-based centrality measures for defining the redundancy of filters. Compared to active filter pruning methods, the proposed passive pruning method gives a similar performance without involving feature maps during pruning. 

In future, we would like to improve the performance of the proposed pruning method by designing better centrality measures and reducing the fine-tuning process overhead further.

\section{ACKNOWLEDGMENT}
\label{sec:ack}

This work was partly supported by a PhD studentship from the Engineering and Physical Sciences Research Council (EPSRC) Doctoral Training Partnership EP/T518050/1 and “AI for Sound (AI4S)” grant EP/T019751/1. For the purpose of open access, the authors have applied a Creative Commons Attribution (CC BY) licence to any Author Accepted Manuscript version arising.


\bibliographystyle{IEEEtran}
\bibliography{refs23.bib}
%
%
%
%
%
%
%
%
%

\end{sloppy}
\end{document}